\documentclass[sigconf]{acmart}

\AtBeginDocument{%
  \providecommand\BibTeX{{%
    \normalfont B\kern-0.5em{\scshape i\kern-0.25em b}\kern-0.8em\TeX}}}

\usepackage{hyperref}
\hypersetup{
    colorlinks=true,
    linkcolor=purple,
    filecolor=magenta,      
    urlcolor=purple,
}
\usepackage{url}            
\usepackage{booktabs}       
\usepackage{amsfonts}       
\usepackage{xcolor}         
\usepackage{makecell}

\newcommand{\ie}{\textit{i}.\textit{e}.}
\newcommand{\eg}{\textit{e}.\textit{g}.}

\usepackage{pifont}
\newcommand{\cmark}{\ding{51}}

\usepackage{arydshln}
\usepackage{wrapfig}
\usepackage{multirow}
\usepackage{amsmath}
\usepackage{graphicx}

\setcopyright{acmlicensed}
\copyrightyear{2024}
\acmYear{2024}
\acmDOI{XXXXXXX.XXXXXXX}

\acmConference[MM'24]{Make sure to enter the correct
  conference title from your rights confirmation email}{October 28 - November 1,
  2024}{Melbourne, Australia.}
\acmISBN{978-1-4503-XXXX-X/18/06}

\begin{document}

\title{Efficient Training for Multilingual Visual Speech Recognition: Pre-training with Discretized Visual Speech Representation}
\renewcommand{\shorttitle}{Efficient Training for Multilingual Visual Speech Recognition}

\author{Minsu Kim$^{*}$}
\affiliation{%
  \institution{KAIST}
  \city{Daejeon}
  \country{South Korea}}
\email{ms.k@kaist.ac.kr}

\author{Jeong Hun Yeo$^{*}$}
\affiliation{%
  \institution{KAIST}
  \city{Daejeon}
  \country{South Korea}}
\email{sedne246@kaist.ac.kr}

\author{Se Jin Park}
\affiliation{%
  \institution{KAIST}
  \city{Daejeon}
  \country{South Korea}}
\email{jinny960812@kaist.ac.kr}

\author{Hyeongseop Rha}
\affiliation{%
  \institution{KAIST}
  \city{Daejeon}
  \country{South Korea}}
\email{ryool_1832@kaist.ac.kr}

\author{Yong Man Ro$^{\dagger}$}
\affiliation{%
  \institution{KAIST}
  \city{Daejeon}
  \country{South Korea}}
\email{ymro@kaist.ac.kr}

\thanks{$^{*}$Both authors have contributed equally to this work. $^\dagger$Corresponding Author.}

\renewcommand{\shortauthors}{Minsu Kim and Jeong Hun Yeo, et al.}

\begin{abstract}
This paper explores sentence-level multilingual Visual Speech Recognition (VSR) that can recognize different languages with a single trained model. As the massive multilingual modeling of visual data requires huge computational costs, we propose a novel training strategy, processing with visual speech units. Motivated by the recent success of the audio speech unit, we propose to use a visual speech unit that can be obtained by discretizing the visual speech features extracted from the self-supervised visual speech model. Through analysis, we verify that the visual speech units mainly contain viseme information while suppressing non-linguistic information. By using the visual speech units as the inputs of our system, we propose to pre-train a VSR model to predict corresponding text outputs on multilingual data constructed by merging several VSR databases. As both the inputs (i.e., visual speech units) and outputs (i.e., text) are discrete, we can greatly improve the training efficiency compared to the standard VSR training. Specifically, the input data size is reduced to 0.016\% of the original video inputs. In order to complement the insufficient visual information in speech recognition, we apply curriculum learning where the inputs of the system begin with audio-visual speech units and gradually change to visual speech units. After pre-training, the model is finetuned on continuous features. We set new state-of-the-art multilingual VSR performances by achieving comparable performances to the previous language-specific VSR models, with a single trained model.
\end{abstract}

\vspace{-0.1cm}
\keywords{Visual Speech Recognition, Lip Reading, Multilingual VSR}

\maketitle
\vspace{-0.1cm}
\section{Introduction}
These days, speech processing technologies have made great progress in diverse applications such as speech recognition \cite{guo2021recent,shim2021understanding,prabhavalkar2023end,hong2023watch}, speech synthesis \cite{wu2022adaspeech,choi2023intelligible,zhang2023speak,jiang2023mega,maiti2023voxtlm}, and speech translation \cite{inaguma2019multilingual,jia2022translatotron,lee2022textless,kim2023many}. Now, it is easy to find a speech processing model that can proficiently handle approximately 100 languages \cite{adams2019massively,radford2023whisper}. However, multilingualism has been mainly explored for audio-based speech processing systems \cite{toshniwal2018multilingual,lux2022low}, while visual-based speech processing systems are still tied to developing monolingual systems \cite{ma2022visual,kim2023lip,yeo2023visual}. There are two reasons for the lagging development of visual-based speech processing systems: 1) The high dimensionality of visual data compared to audio puts a challenge in training a large-scale model with massive multilingual data. Compared to the same length of audio, visual data requires about six times larger bits \cite{kim2023practical} in a standard visual speech recognition process \cite{ma2021end}. Moreover, the requirement of encoding spatial information using two-dimensional convolutions also increases the computation costs of visual speech processing compared to its counterpart. 2) The low quantity of labeled data in visual speech processing systems presents a formidable obstacle to technology development. In contrast to the tremendous amount of publicly available audio-text data \cite{pratap2020mls}, a very limited number of video-text data are available, especially for non-English \cite{kim2023lip}.

In this paper, we explore the multilingualism of visual speech processing, especially in speech recognition \cite{assael2016lipnet,petridis2016deep,chung2017lip,ma2021towards,ma2022training}. Hence, our objective is to devise a multilingual Visual Speech Recognition (VSR) method that can recognize different languages with a single trained model. In order to mitigate the challenges in visual speech processing, we propose a novel strategy, processing with visual speech units. The audio speech unit \cite{lakhotia2021generative} is a discretized representation of an extracted speech feature from a self-supervised speech model \cite{baevski2020wav2vec,hsu2021hubert}. It contains phonemic content \cite{sicherman2023analysing} while suppressing the other speech characteristics ($\eg$, speaker information) and can be employed as pseudo text. As it is the discretized signal of the original signal, the data size can be significantly reduced \cite{chang2023exploration,kim2023practical}. Motivated by this, we propose to employ visual speech units, the quantized representation of the visual speech feature, in training multilingual VSR. As a result, one video frame having 61,952 bits (\ie, based on a grayscale image with 88 $\times$ 88 size) can be expressed with one visual speech unit which can be represented with just 10 bits. With the huge data size reduction, 0.016\% compared to the original, we can boost the training more than 10 times faster than the standard VSR training. Through analysis, we validate that the visual speech unit contains viseme information, the visual counterpart of phoneme, while suppressing non-linguistic characteristics. Hence, enabling visual speech modeling even by using the visual speech units.

Specifically, we employ AV-HuBERT \cite{shi2021learning}, a self-supervised visual speech model, to extract our visual speech units. We newly train the AV-HuBERT model on 5,512 hours of multilingual audio-visual data composed of nine languages, given that the original AV-HuBERT was trained solely on English audio-visual data. To differentiate between the English-only trained and multilingual-trained AV-HuBERT models, we denote the latter as mAV-HuBERT. With the mAV-HuBERT, we can correctly capture the multilingual viseme information into our visual speech unit. Then, we propose to pre-train an encoder-decoder model by setting its inputs with the visual speech units and the outputs with the corresponding text, forming a unit-to-unit translation framework (\ie, translation between discrete tokens) \cite{kim2023many}. Moreover, inspired by the recent successes of VSR that leverage audio modal information to complement limited visual data \cite{zhao2020hearing,ren2021learning,kim2021cromm,shi2021learning,haliassos2022jointly}, we propose to use curriculum learning with a gradual increase in task difficulty using audio modality. Concretely, our unit-to-unit pre-training is initiated with audio-visual inputs and then gradually changed to visual inputs. With this curriculum learning, the model can find the optimization points stable and achieve higher performance with the complementary multi-modal information. To mitigate the comparatively small amount of public visual-text paired data, we utilize the recently proposed automatic labels of \cite{ma2023auto} and \cite{yeo2023visual} where the text labels are obtained by their automatic labeling processes. Finally, the pre-trained model is finetuned with continuous input features to maximize the VSR performances.

\begin{figure*}[t]
	\centering
	\centerline{\includegraphics[width=15cm]{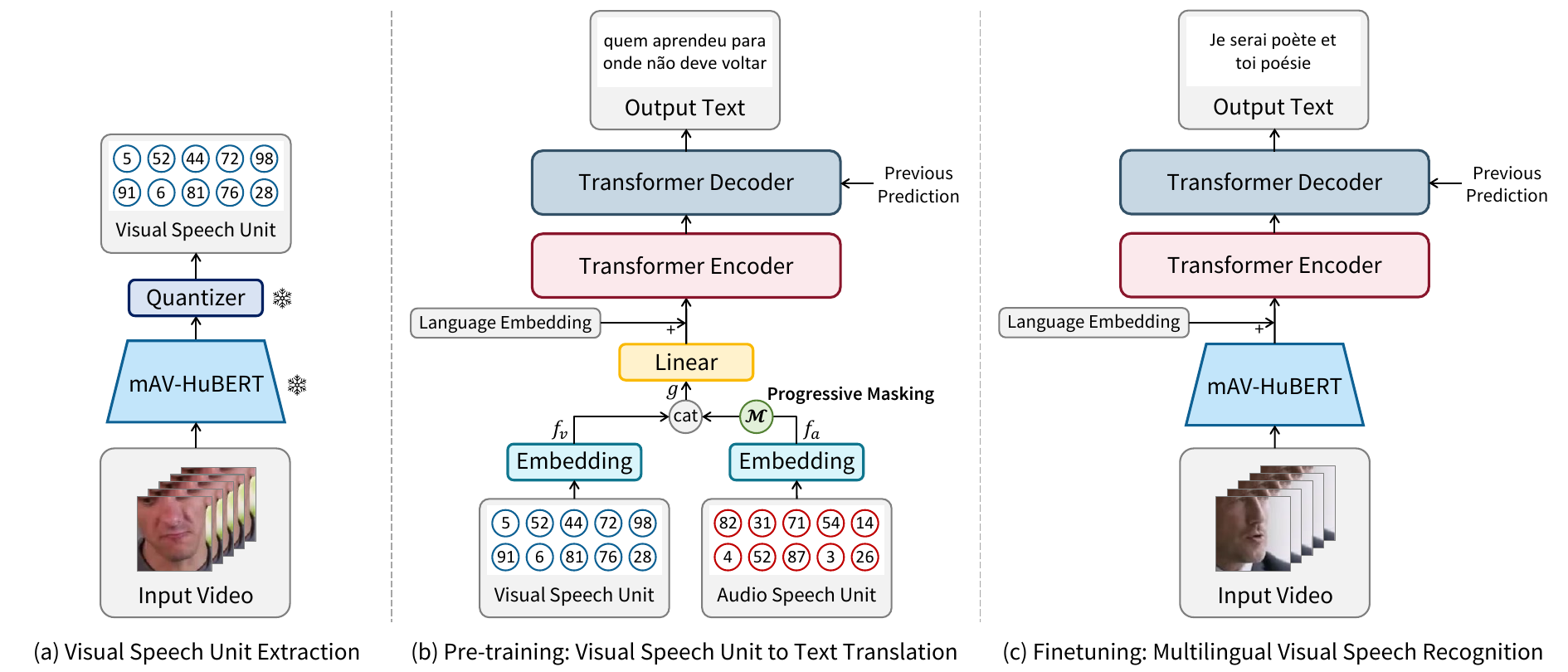}}
    \vspace{-0.2cm}
	\caption{Illustration of the proposed multilingual VSR framework. (a) Visual speech units are obtained by quantizing the visual speech features extracted from mAV-HuBERT. (b) The Transformer encoder-decoder model is pre-trained with discrete inputs and outputs. The task difficulty is gradually increased by using progressive masking $\mathcal{M}$. Pre-training commences with audio-visual speech units as inputs, and these inputs gradually transition to visual speech units through progressive masking of the audio speech units. (c) After pre-training the model with discrete inputs and outputs, it is finetuned with continuous features to boost the VSR performances.
	}
	\label{fig:1}
\vspace{-0.3cm}
\end{figure*}

The major contributions can be summarized as follows: 
\begin{itemize}
    \item To the best of our knowledge, this is the first work exploring sentence-level multilingual VSR with a single trained model. The proposed VSR model can recognize five different languages using a single model, whereas previous methods required training separate models for each language.
    \item We propose to employ visual speech units as inputs to pre-train the multilingual VSR model, thereby establishing discrete inputs and outputs (\ie, text). With this, we can drastically reduce the computational costs and accelerate the pre-training time by about 10 times compared to the standard training.
    \item Through analysis, we verify that the visual speech unit mainly holds the viseme information while suppressing non-linguistic features, enabling VSR training even by employing discretized inputs. 
    \item We set new state-of-the-art multilingual VSR performances by achieving comparable performances with the multiple previous monolingual VSR methods.
\end{itemize}

\vspace{-0.1cm}
\section{Related Work}
\subsection{Visual Speech Recognition (VSR)}
Visual Speech Recognition (VSR) aims to predict the spoken words from silent lip movements video. Early works \cite{chung2017lrw,stafylakis2017combining,petridis2017lstm,petridis2018end} focused on word-level VSR by using CNN \cite{he2016deep} and the RNN \cite{chung2014empirical, hochreiter1997long}. Large-scale lip-reading sentence datasets \cite{chung2017lrs2, afouras2018lrs3} have boosted the development of sentence-level VSR. By employing Transformer \cite{vaswani2017attention} architecture, \cite{afouras2018deep} proposed a powerful sentence-level end-to-end VSR model. Moreover, the integration of the hybrid CTC/Attention objective \cite{watanabe2017hybrid,petridis2018audio} into VSR, greatly improved the recognition performances.
Recent VSR technologies \cite{ma2021end,prajwal2022sub,chang2023conformers,ma2023auto} also employed transformer-variant architectures and improved the VSR performances. For advanced training strategies, many researchers try to reduce the gap between visual and audio modalities. They \cite{zhao2020hearing,afouras2020asr,ren2021learning, ma2021towards,kim2021cromm,kim2022distinguishing,yeo2023multi} studied how to effectively transfer audio knowledge into the VSR model by using knowledge distillation \cite{hinton2015distilling} and memory network \cite{weston2014memory}. However, these previous VSR approaches have mainly developed for high-resource languages, English and Mandarin \cite{luo2020synchronous}. VSR for different languages, especially low VSR resource languages, has only been addressed recently \cite{ma2022visual, zinonos2023learning, kim2023lip, yeo2023visual}. In particular, a recent approach \cite{yeo2023visual} proposed the labeled data for low VSR resource languages using automatic labeling processes. 

This paper is the first work exploring sentence-level multilingual VSR with a single model. To mitigate the huge computational costs in training the multilingual VSR model, we propose to pre-train the model with discrete inputs and outputs by using visual speech units. To complement the low amount of video-text data, we bring the automatic labels of \cite{ma2023auto,yeo2023visual} and propose curriculum learning that utilizes audio modality to provide rich speech information.

\vspace{-0.1cm}
\subsection{Audio speech unit}
Audio speech unit \cite{lakhotia2021generative} is the discretized speech representation of self-supervised speech models such as HuBERT \cite{hsu2021hubert}, Wav2Vec2.0 \cite{baevski2020wav2vec}, and WavLM \cite{chen2022wavlm}. It is possible to suppress non-linguistic features and mainly keep the linguistic contents by selecting proper layers to extract the speech features \cite{lakhotia2021generative,polyak2021speech}. By using the speech unit as pseudo text, Textless Natural Language Processing becomes possible \cite{lee2022textless,huang2022transpeech,popuri2022enhanced,kim2023many,nguyen2023generative}. Moreover, speech units have promising potential to be used in multi-modal processing as they greatly reduce the data size \cite{chang2023exploration,park2023storage,kim2023practical}.

Motivated by this, we propose to employ a visual speech unit which is the quantized representation of visual speech features extracted from the self-supervised visual speech model \cite{shi2021learning}. We analyze the characteristics of visual speech units and show that the visual speech unit contains mainly viseme information while suppressing the other characteristics.

\vspace{-0.1cm}
\section{Method}
\label{sec:3}
The objective of this paper is to develop a multilingual VSR model, so that multiple languages can be recognized by using a single trained model. To mitigate the large computational costs in developing visual speech processing systems, we propose visual speech units which are discretized representations of visual speech features encoded by a self-supervised speech model.

\vspace{-0.1cm}
\subsection{Multilingual Visual Speech Unit Extraction}
Audio speech units \cite{lakhotia2021generative} can be obtained by clustering the speech features of a self-supervised speech model such as HuBERT \cite{hsu2021hubert}. Analogous to audio speech units, we propose to employ visual speech units, which can be obtained by quantizing the visual speech features derived from a pre-trained visual speech model. In order to get visual speech units, we choose AV-HuBERT \cite{shi2021learning} for the self-supervised visual speech model, which is well-known for its discriminative visual speech features. However, AV-HuBERT is pre-trained on English-only audio-visual data, which deviates from our primary objective of achieving multilingualism. Hence, we initially train a multilingual variant of AV-HuBERT (mAV-HuBERT) to ensure the accurate incorporation of multilingual viseme information into the final visual speech units. To this end, we train the model on 5,512 hours of multilingual dataset composed of 9 languages (En, Es, It, Fr, Pt, De, Ru, Ar, El) by merging LRS2 \cite{chung2017lrs2}, LRS3 \cite{afouras2018lrs3}, VoxCeleb2 \cite{chung2018voxceleb2}, and AVSpeech \cite{ephrat2018avspeech}. As VoxCeleb2 and AVSpeech do not have language identities, we obtain the language identity of each utterance by using a pre-trained language identifier of Whisper \cite{radford2023whisper}, to select the data. For the prediction target of the masked prediction of mAV-HuBERT, we use clusters of speech features obtained from a pre-trained multilingual HuBERT \cite{hsu2021hubert,lee2022textless}. We use the target size of 1,000 and train the model for 350k steps with one iteration. Through analysis we confirm that mAV-HuBERT is more suitable for multilingual speech modeling than the English-only trained AV-HuBERT in Sec. \ref{sec:4.3.1}.

With the pre-trained mAV-HuBERT, we extract the visual speech unit by clustering (\ie, quantizing) the output visual speech features, as shown in Fig.~\ref{fig:1}(a). Please note that AV-HuBERT can extract both audio-only features and visual-only features through its modality dropout. Hence, we only use the visual inputs to extract the visual speech units. For the token size of the visual speech unit, we use 1,000 so that each visual speech unit can be represented with just 10 bits. Please note that one video frame having grayscale and 88 $\times$ 88 size (\ie, the standard for visual speech recognition) requires 61,952 bits \cite{ma2021end,shi2021learning}. Therefore, we can reduce the data size to 0.016\% compared to the raw visual inputs, which enables us to greatly increase the training batch size and accelerate the training speed by removing the visual front-end (\eg, 2D CNNs). We analyze the efficiency of visual speech units by comparing them with the standard raw inputs in Sec. \ref{sec:4.3.2}.

\begin{table*}[t]
\renewcommand{\arraystretch}{1.3}
\renewcommand{\tabcolsep}{2.5mm}
\centering
\caption{Dataset used for pre-training mAV-HuBERT and training the multilingual VSR model. Audio-visual data is utilized for mAV-HuBERT, while video-text data is utilized for the multilingual VSR model (\ie, for both pre-training and finetuning).}
\resizebox{0.8\linewidth}{!}{  
    \begin{tabular}{ccc}
    \Xhline{3\arrayrulewidth}
    \multicolumn{3}{c}{\textbf{Train Data for mAV-HuBERT}} \\ \hline
     \textbf{Datasets} & \makecell{\textbf{Number of Video} \\ \textbf{/ Hours}} & \textbf{Languages}\\ \hline
     \makecell{LRS2} & 142,157 / 223 & En \\
    \makecell{LRS3} & 150,498 / 433 & En \\
    \makecell{mTEDx} & 181,034 / 285 & Es, Fr, It, Pt \\
    \makecell{VoxCeleb2}  & 834,375 / 1,739 & En, Es, It, Fr, Pt, De, Ru, Ar, El \\
    \makecell{AVSpeech} & 1,575,755 / 2,832 & En, Es, It, Fr, Pt, De, Ru, Ar, El \\ \hline
    \makecell{Total} & 2,883,819 / 5,512 & En, Es, It, Fr, Pt, De, Ru, Ar, El \\ 
    \Xhline{3\arrayrulewidth}
    \end{tabular}
    \hspace{0.1em}
    \begin{tabular}{ccc}
    \Xhline{3\arrayrulewidth}
    \multicolumn{3}{c}{\textbf{Train Data for multilingual VSR}} \\ \hline
     \textbf{Datasets} & \makecell{\textbf{Number of Video} \\ \textbf{/ Hours}} & \textbf{Languages}\\
    \hline
    - & - & - \\
    \makecell{LRS3} & 150,498 / 433 & En \\
    \makecell{mTEDx} & 181,034 / 285 & Es, Fr, It, Pt \\
    \makecell{VoxCeleb2} & 742,147 / 1,539  & En, Es, It, Fr, Pt \\
    \makecell{AVSpeech} & 1,272,065 / 2,288 & En, Es, It, Fr, Pt \\ 
    \hline
    \makecell{Total} & 2,505,858 / 4,545 & En, Es, It, Fr, Pt \\ 
    \Xhline{3\arrayrulewidth}
    \end{tabular}}
\label{table:1}
\vspace{-0.3cm}
\end{table*}

\vspace{-0.1cm}
\subsection{Pre-training: Visual Speech Unit to Text Translation}
By representing all training video data into visual speech units, we can significantly reduce the data size, enabling efficient training of a VSR model on large-scale multilingual data. Based on this key concept, we propose to pre-train our model to predict text by setting the inputs with visual speech units. Therefore, now the inputs and outputs are both discrete, which is illustrated in Fig.~\ref{fig:1}(b). As the visual speech units mainly contain linguistic information, we can pre-train the model to construct the knowledge of visual speech modeling even by using discrete inputs. We analyze the information contained in the visual speech units and validate how it can be worked, in Sec. \ref{sec:4.3.3}.

Nevertheless, translating visual speech units directly into output text from scratch is challenging for the model in identifying optimal solutions. Since visual information contains scarce speech information \cite{zhao2020hearing,kim2021cromm,ren2021learning} compared to audio, training the model directly to perform visual-to-text conversion might be hard to find the solution. To mitigate this, we bring the motivation from the recent success of VSR, which utilizes auxiliary audio information during training \cite{afouras2020asr,zhao2020hearing,shi2021learning,ren2021learning,kim2022distinguishing,ma2022visual,yeo2023multi}. Specifically, we initiate the pre-training with audio-visual speech units where both audio speech units and visual speech units are utilized as inputs, similar to \cite{shi2021learning,djilali2023lip2vec}. Then, we gradually masked out the audio speech units as the pre-training progressed, resulting in the final training stage exclusively utilizing visual speech units as inputs. Therefore, the model can easily find the optimization points through this curriculum learning, with the aid of complementary audio speech information. Concretely, the embeddings of the visual speech units $f_v\in\mathbb{R}^{T\times D}$ and audio speech units $f_a\in\mathbb{R}^{T\times D}$ are concatenated as, $g=\mathcal{M}(f_a)\oplus f_v$, where $T$ is the sequence length, $D$ is the dimension of embedding, $\oplus$ represents concatenation operation in the embedding dimension, and $g\in\mathbb{R}^{T\times 2D}$ is the concatenated feature. $\mathcal{M}(\cdot)$ is a masking function that randomly masks out $p\%$ of frames from the input sequence. We progressively increase the $p$ from 0 to 100 as the pre-training progresses so that the transition from audio-visual inputs to visual inputs can be made. The effectiveness of this curriculum learning using input transition from audio-visual to visual can be found in Sec. \ref{sec:4.3.4}.

Then, we reduce the embedding dimension of the concatenated feature $g$ using a linear layer. Here, we provide the language information by providing the language embedding which will be added to the feature, following \cite{conneau2019cross}. Finally, through the Transformer encoder-decoder architecture, we translate the visual inputs into the output text in an auto-regressive manner. The objective function of our learning problem can be represented as follows,

\begin{align}
    \mathcal{L}=- \sum_{s=1}^S \log P(y_s|\mathbf{X},y_{<s}),
\end{align}

where $y_s$ is the text annotation for current step $s$ and $y_{<s}$ is the previous outputs, $\mathbf{X}$ is the input speech units, and $S$ is the length of the text. For multilingual training, we use five languages (En, Pt, Es, Fr, It) by merging LRS3 \cite{afouras2018lrs3}, mTEDx \cite{elizabeth2021mtedx}, automatic labels for En of \cite{ma2023auto}, and automatic labels for Pt, Es, Fr, and It of \cite{yeo2023visual}, forming 4,545 hours of data. We summarize the dataset statistics in Table \ref{table:1}.

\vspace{-0.1cm}
\subsection{Finetuning: Multilingual Visual Speech Recognition}
Even though we can directly perform multilingual VSR with the pre-trained model with discrete inputs, it is hard to outperform the model using continuous features. This is expected, as information is lost during the quantization process. However, as the pre-trained model has already learned how to model the multilingual visual pronunciation and generate languages, the finetuning with the continuous features is straightforward and more effective than direct training of the multilingual VSR model from scratch. For finetuning, we detach the unit embedding and linear layers, and attach the pre-trained mAV-HuBERT, as illustrated in Fig.~\ref{fig:1}(c). The model is trained end-to-end with the same objective function and training data as the pre-training. In Sec. \ref{sec:4.3.5}, we analyze the performances of multilingual VSR using the single proposed model and multiple previous monolingual models.

\vspace{-0.1cm}
\section{Experiment}
\label{sec:4}
\subsection{Dataset}
{\bf Lip Reading Sentences 2 (LRS2)} \cite{chung2017lrs2} is one of the largest English datasets for VSR. The dataset consists of 223 hours of training data collected from British TV shows.

{\bf Lip Reading Sentences 3 (LRS3)} \cite{afouras2018lrs3} is a popular English VSR database. It has about 430 hours of video, and each video clip is collected from TED and TEDx. We evaluate the English VSR performances on LRS3.

{\bf Multilingual TEDx (mTEDx)} \cite{elizabeth2021mtedx} is a multilingual dataset originally proposed for speech recognition and translation. The dataset provides 8 languages collected from TEDx talks. As the dataset also provides the video links of original talks, we download the video online. We remove the unavailable videos for VSR by referring to \cite{ma2022visual}. We use four languages, Spanish (Es), Italian (It), French (Fr), and Portuguese (Pt) for training and evaluating the developed VSR model, following \cite{ma2022visual,kim2023lip,yeo2023visual}.

{\bf VoxCeleb2} \cite{chung2018voxceleb2} 
is a multilingual audio-visual dataset for speaker recognition \cite{jung2022large}. This dataset has over 1 million utterances and contains 6,112 celebrities. For mAV-HuBERT training, we use the data corresponding to the following 9 languages, English (En), Spanish (Es), Italian (It), French (Fr), Portuguese (Pt), German (De), Russian (Ru), Arabic (Ar), and Greek (El), by identifying the language identity using Whisper \cite{radford2023whisper}. For VSR training, as the dataset does not provide text annotations, we use the automatic labels of \cite{ma2023auto} for En, and \cite{yeo2023visual} for Es, It, Fr, and Pt.

{\bf Audio Visual Speech Dataset (AVSpeech)} \cite{ephrat2018avspeech} is a large-scale audio-visual speech dataset. AVSpeech contains roughly 290k YouTube videos, and the total duration of these videos is 4700 hours. Since the dataset does not provide text annotations, we use the same strategy with the VoxCeleb2, using 9 languages for mAV-HuBERT and automatic labels for VSR.

\vspace{-0.1cm}
\subsection{Implementation Details}
\textbf{Preprocessing.} The video is resampled to 25 fps. We detect the facial landmarks using RetinaFace \cite{deng2020retinaface}, crop mouth regions using $96 \times 96$ sizes of bounding box, and convert them into grayscale. For data augmentation, we randomly crop the video into $88 \times 88$ and horizontally flip it during training. The audio is resampled to 16kHz. In order to obtain the audio speech unit, we feed the audio into a multilingual trained HuBERT \cite{hsu2021hubert,lee2022textless} and cluster the extracted features with 1,000 token size. Finally, the audio speech unit is resampled to 25 fps to align with the sampling rate of the visual inputs. For the text, we construct one multilingual dictionary with 1,000 subword units by using SentencePiece tokenizer \cite{kudo2018sentencepiece}.

\textbf{Architecture.} The mAV-HuBERT has the same architecture as the AV-HuBERT \cite{shi2021learning} large configuration and only differs in pre-training data. For visual speech unit-to-text translation, the model is composed of two unit embedding layers, one linear layer, one language embedding layer, 6 transformer encoders, and 6 transformer decoders. Each unit embedding layer embeds 1,000 tokens into 1,024 dimensions and the linear layer reduces the concatenated 2,048 dimensions into 1,024. The language token is embedded into a 1,024-dimensional feature by the language embedding layer. Each transformer layer has an embedding dimension of 1024, a feed-forward dimension of 4096, and 16 heads.

\begin{table}[t]
\renewcommand{\arraystretch}{1.6}
\renewcommand{\tabcolsep}{1.2mm}
\centering
\caption{Comparisons between English AV-HuBERT and the multilingual AV-HuBERT in multilingual VSR. English (En) is validated on LRS3 and other languages (Es, It, Fr, Pt) are validated on mTEDx databases.}
  \resizebox{0.999\linewidth}{!}{
  \begin{tabular}{ccccccc}
    \Xhline{3\arrayrulewidth}
    \multirow{2}{*}{\textbf{Method}} & \multirow{2}{*}{\textbf{Finetune Datasets}} & \multicolumn{5}{c}{\textbf{WER (\%)}}  \\
    \cline{3-7}
    & & \textbf{En} & \textbf{Es} & \textbf{It} & \textbf{Fr} & \textbf{Pt} \\
    \hline
    AV-HuBERT \cite{shi2021learning} & LRS3, mTEDx & \textbf{28.0} & 75.9 & 74.0 & 75.5 & 79.6 \\ \hdashline
    \textbf{mAV-HuBERT (Ours)} & LRS3, mTEDx & 33.7 & \textbf{54.3} & \textbf{59.1} & \textbf{63.0} & \textbf{58.8} \\ 
    \Xhline{3\arrayrulewidth}
  \end{tabular}}
\label{table:2}
\vspace{-0.5cm}
\end{table}
\begin{table}[t]
\renewcommand{\arraystretch}{1.2}
\renewcommand{\tabcolsep}{1.0mm}
\centering
\caption{Efficiency comparisons between previous VSR method and the proposed method. All numbers are measured using the same CPU and GPU (RTX 3090 24GB) environments. Test Acc means the subword-level prediction accuracy without beam search decoding.}
  \resizebox{0.999\linewidth}{!}{
  \begin{tabular}{cccccc}
    \Xhline{3\arrayrulewidth}
    \textbf{Method} & \textbf{Input} & \makecell{\textbf{Batch Size} \\ \textbf{(Tot. Frames)}} & \makecell{\textbf{Train Iter.} \\ \textbf{Time (sec)}} & \makecell{\textbf{Tot. Train} \\ \textbf{Time (hrs)}} & \makecell{\textbf{Test} \\ \textbf{Acc (\%)}}  \\
    \hline
    Standard VSR & Video & 1,000 & 1.58 & 52.5 & 82.1\\ \hdashline
    \textbf{Pre-training} & Speech Unit & 6,000 & 0.88 & 6.6 &  72.2  \\
    \textbf{Finetuning} & Video & 1,000 & 1.68 & 34.9 & 82.2\\ 
    \Xhline{3\arrayrulewidth}
  \end{tabular}}
\label{table:3}
\vspace{-0.6cm}
\end{table}

\textbf{Training.} For training mAV-HuBERT, we follow the original AV-HuBERT \cite{shi2021learning} and train it with the masked prediction task. For the prediction target, we use 1,000 clusters extracted from multilingual trained HuBERT \cite{hsu2021hubert,lee2022textless}. We train the model for 350k steps using 64 3090 RTX GPUs. 
For pre-training the proposed model (\ie, visual speech unit-to-text translation), we train the model for 11 epochs with a tri-stage learning rate scheduler and 32 GPUs. For the progressive masking $\mathcal{M(\cdot)}$, we set $p$ as 0 for the first 10\% of training. From 10\% to 70\% of training, we linearly increase the masking ratio $p$ from 0 to 100. After 70\% of training, $p$ is set to 100 so that only visual speech units are used. 
We finetune the model with the continuous features for 8 epochs using 32 GPUs. For all experiments, the Adam optimizer \cite{kingma2014adam} is used. For beam search decoding, we use a beam width chosen from \{20, 25, 30, 35\} and a length penalty of 0. The detailed training configuration can be found in the supplementary.

\vspace{-0.1cm}
\subsection{Experimental Results}
\subsubsection{Effectiveness of mAV-HuBERT in Modeling Multilingual Speech}
\label{sec:4.3.1}
Before extracting the visual speech units by using SSL visual speech models, we need to confirm which model is best suitable for our multilingual VSR purposes. Therefore, we compare the performances of English-trained AV-HuBERT and multilingual-trained AV-HuBERT (mAV-HuBERT). To this end, we finetune the pre-trained mAV-HuBERT and AV-HuBERT on 5 languages (\ie, En, Es, It, Fr, Pt) by merging LRS3 and mTEDx databases. The multilingual VSR performances of the two models are shown in Table \ref{table:2}. The results show that when performing multilingual VSR with a single model, the mAV-HuBERT is much more effective than the English-only trained AV-HuBERT. Specifically, for Es, It, Fr, and Pt, the mAV-HuBERT outperforms the AV-HuBERT with large margins of over 10\% WERs.
For English, English-only trained AV-HuBERT achieves better performance, and this tendency is also observed in \citet{radford2023whisper}; Increasing language diversity without scaling the model size, the performances for major languages can be lower than the monolingual model.
As the mAV-HuBERT shows better performances in modeling multilingual visual speech, we extract the visual speech unit by using the mAV-HuBERT.

\begin{figure}[t]
	\centering
	\centerline{\includegraphics[width=8.7cm]{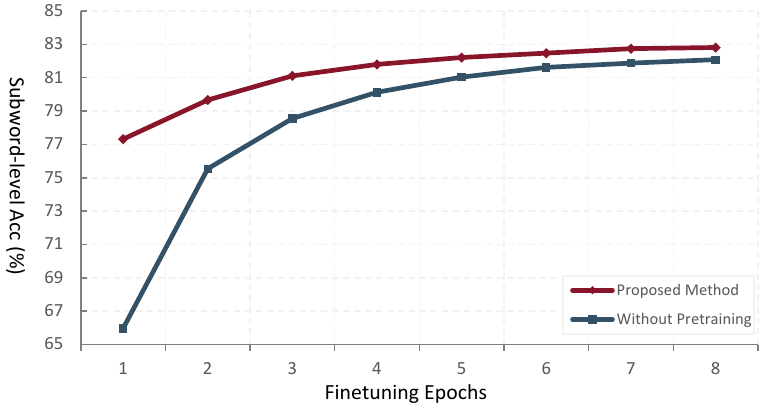}}
\vspace{-0.1cm}
\caption{Finetuning efficiency comparison between the proposed pre-training scheme and without the pre-training.}
\label{fig:2}
\vspace{-0.3cm}
\end{figure}

\begin{figure*}[t]
	\centering
	\centerline{\includegraphics[width=14cm]{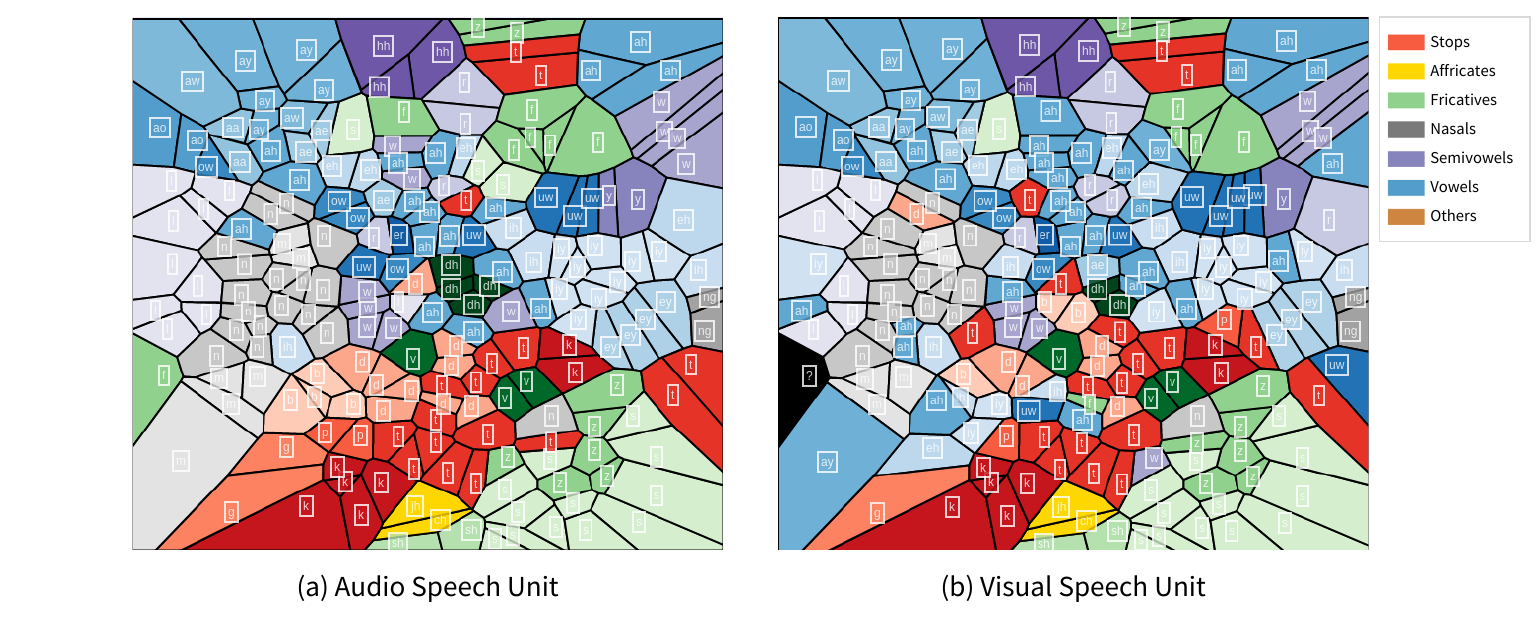}}
    \vspace{-0.3cm}
	\caption{Visualization of speech units. Each boundary represents a single unit and the same color represents the same phoneme or phoneme family. (a) Audio speech unit. (b) Visual speech unit.}
	\label{fig:3}
\vspace{-0.3cm}
\end{figure*}

\begin{table*}[t]
  \renewcommand{\arraystretch}{1.5}
  \renewcommand{\tabcolsep}{2.5mm}
\centering
\caption{Speaker verification results (EER) comparisons using different input representations.}
  \resizebox{0.7\linewidth}{!}{
  \begin{tabular}{ccccc}
    \Xhline{3\arrayrulewidth}
    \textbf{Raw audio} & \textbf{Audio Feature} & \textbf{Visual Feature} & \textbf{Audio speech unit} & \textbf{Visual speech unit} \\ \hline
    2.38\% & 14.96\% & 19.42\% & 28.84\% & 32.74\% \\
    \Xhline{3\arrayrulewidth}
  \end{tabular}}
  \label{table:4}
\vspace{-0.3cm}
\end{table*}

\subsubsection{Efficiency Comparison between Visual Speech Units and Raw Inputs}
\label{sec:4.3.2}
To confirm the efficiency of using visual speech units instead of raw videos as inputs, we compare the batch size, training iteration time, and total training time between the standard VSR method that uses raw video as inputs and the proposed method. Table \ref{table:3} shows the comparison results. By using the visual speech units as inputs in pre-training, we can increase the batch size sixfold and reduce the training iteration time by approximately half. Consequently, we can expedite training by a factor of about 12 compared to previous VSR training methods. The total training time for pre-training the model for 11 epochs amounts to 6.6 hours, whereas the conventional VSR model requires 52.5 hours for 8 epochs of training. Furthermore, when fine-tuning the pre-trained model, we can achieve better performance compared to the standard VSR model, with just 5 epochs of fine-tuning taking 34.9 hours. As a result, we can greatly boost training time even if considering both the pre-training and finetuning stages, compared to the standard VSR training. To see this more intuitive, we examine the subword-level accuracy changes during fine-tuning, comparing the proposed method with the model without pre-training. Fig. \ref{fig:2} shows the learning curve of the two models. After the proposed pre-training, the finetuning on the continuous features is effective so that we can achieve better performance even with much fewer epochs than the model without pre-training. It's worth noting that we can significantly reduce the pre-training time with the proposed method; 1 epoch requires just 0.6 hours in pre-training, whereas the standard VSR training demands 6.6 hours for 1 epoch.

\subsubsection{Analyzing Visual Speech Units}
\label{sec:4.3.3}
\begin{figure*}[t]
	\centering
	\centerline{\includegraphics[width=12cm]{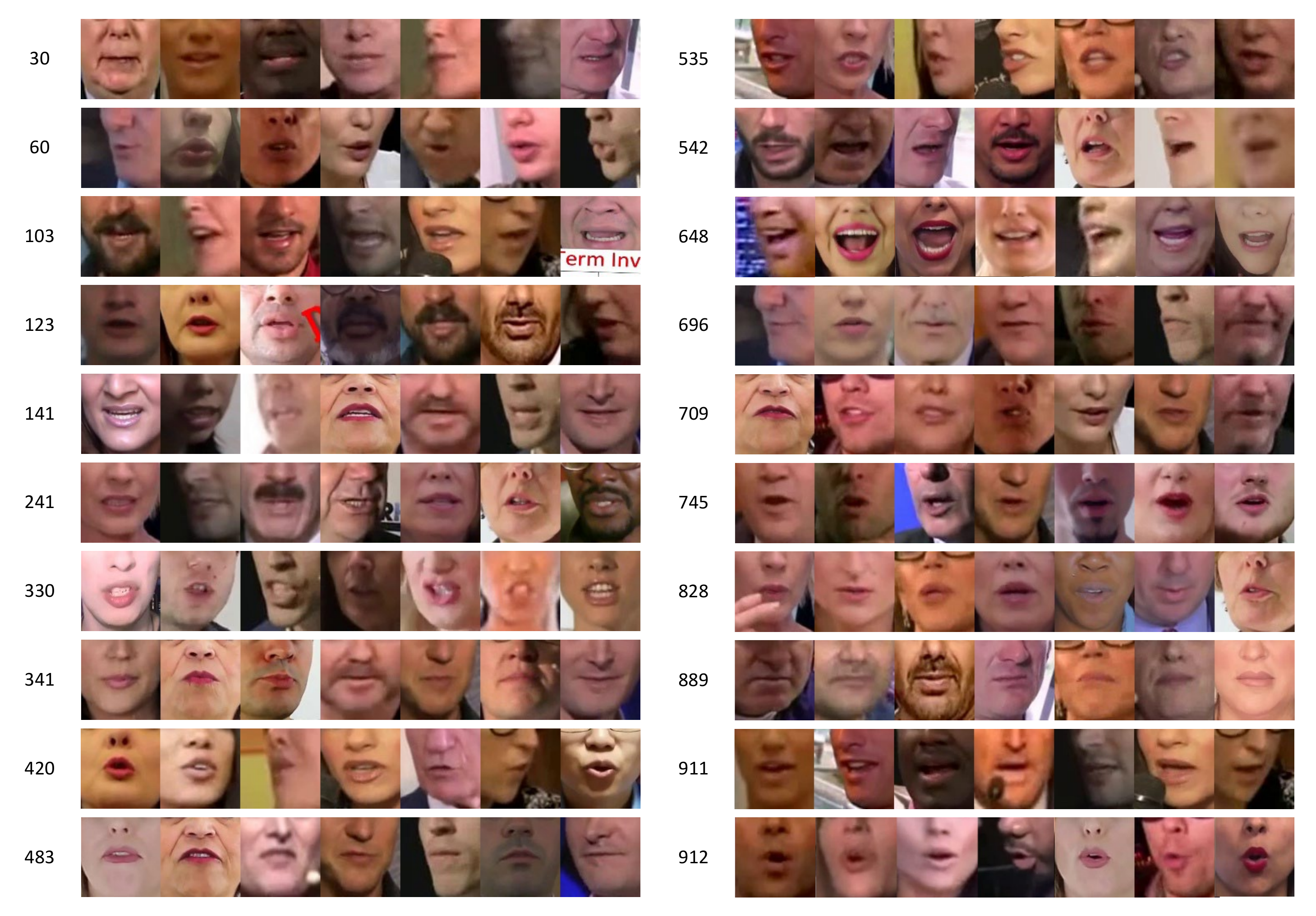}}
	\caption{Example frames corresponding to each visual speech unit. The numbers indicate the classes of visual speech units.}
	\label{fig:4}
\vspace{-0.3cm}
\end{figure*}

To understand which information is held by the visual speech units, we analyze them using 1) phoneme mapping visualization \cite{sicherman2023analysing} and 2) speaker recognition. Firstly, following \cite{sicherman2023analysing}, we visualize the phonetic information of the audio speech unit and visual speech unit which are obtained from mAV-HuBERT. Therefore, only audio inputs are used to extract audio speech units and video inputs for visual speech units. We set the cluster centroid as each row of the weight of the final layer of mAV-HuBERT (\ie, classifier for 1,000 units), so that the same boundary can be obtained for different modalities. Fig.~\ref{fig:3} displays the visualization results of the 200 units out of 1,000 that appeared most frequently. We can confirm that each unit in both the audio speech units and visual speech units contains distinctive linguistic information (\ie, phoneme or viseme). By comparing the audio speech unit and visual speech unit, we can find that the homophenes which refer to the different pronunciations having the same lip movement (\eg, \underline{n}ame, \underline{t}ame, \underline{d}ame) are confusingly represented in visual speech units \cite{kim2022distinguishing}. For example, some units representing `n' in audio speech units are changed to `d' or `t' in visual speech units. Moreover, we can find that compared to audio speech units, more visual speech units are pointing to the vowel (\ie, blue-colored area), which shows the ambiguity of lip movements compared to the audio modality. These are the natural results reflecting the characteristics of different speech modalities, and should not be taken as an indication of the inadequacy of the visual modality in speech modeling. In addition, we visualize some video frames corresponding to each visual speech unit in Fig.~\ref{fig:4}, illustrating that similar lip movements are consistently mapped to the same index regardless of pose variations and speakers. For example, the 648-th visual speech unit represents lip frames related to the viseme `a', and the 912-th visual speech unit corresponds to the viseme `o'. Through the visualization, we can confirm that our visual speech units contain the linguistic information, the viseme, which enables us to pre-train the model to build the knowledge of visual speech modeling.

Secondly, we analyze how much degree the speech units contain non-linguistic information through speaker recognition. To this end, we use raw audio speech, audio/visual speech features extracted from mAV-HuBERT, and audio/visual speech units as inputs to perform the speaker recognition. For the model, we utilize a pre-trained speaker recognition model of \cite{desplanques2020ecapa} and train the model with different inputs. To match the input size with the model, we use one additional embedding layer for both audio speech units and visual speech units. When we use the speech features from mAV-HuBERT, we utilize one additional linear layer for both audio speech features and visual speech features. For training, we utilize data less than 20 seconds in VoxCeleb2 \cite{chung2018voxceleb2} dev set. The speaker verification is performed on the test data of VoxCeleb2 having less than 20 seconds, where each test sample is assigned one positive sample and one negative sample. Therefore, 27,816 positive pairs and negative pairs are utilized respectively, thus random prediction yields 50\% EER.

The speaker verification results (Equal Error Rate; EER) are shown in Table \ref{table:4}. When we use raw audio as input, the system can almost perfectly distinguish the input speakers with 2.38\% EER. When we use the features of the pre-trained mAV-HuBERT, the verification performance is dropped to 14.96\% and 19.42\% on audio and visual modalities, respectively. This shows that the masked prediction of AV-HuBERT \cite{shi2021learning} forces the model to learn linguistic information while somewhat discarding the speaker information. Finally, when we discretize the speech features of mAV-HuBERT and obtain the speech units, we can greatly suppress the speaker information. The performance is dropped to 28.84\% by using the audio speech unit and 32.74\% by using the visual speech unit. Through the above two experiments, we can confirm that the quantization process averages out the speaker effects (\ie, non-linguistic information) in the speech representations while maintaining the content. This enables us to build a speech representation model by using speech units as input.

\begin{table}[t]
\renewcommand{\arraystretch}{1.2}
\renewcommand{\tabcolsep}{3.5mm}
\centering
\caption{The effectiveness of each proposed component (WER, \%). By substituting each component from the full model, we evaluate their effectiveness in multilingual VSR.}
  \resizebox{0.9\linewidth}{!}{
  \begin{tabular}{lccccc}
    \Xhline{3\arrayrulewidth}
    \textbf{Method} & \textbf{En} & \textbf{Es} & \textbf{It} & \textbf{Fr} & \textbf{Pt} \\ \hline
    \textbf{Proposed Method} & 24.4 & 42.7 & 49.7 & 55.2 & 50.6 \\ \hdashline
    $-$ Unit Pretraining & 24.3 & 45.1 & 52.1 & 56.2 & 51.1 \\
    $-$ Curriculum & 25.2 & 46.3 & 52.5 & 55.8 & 50.6 \\
    $-$ Finetuning & 37.5 & 68.1 & 70.3 & 76.3 & 74.0 \\
    \Xhline{3\arrayrulewidth}
  \end{tabular}}
\label{table:5}
\vspace{-0.3cm}
\end{table}

\subsubsection{Effectiveness of Each Proposed Component}
\label{sec:4.3.4}
To confirm the effectiveness of each proposed component, we substitute each component from the proposed method. The ablation results are shown in Table \ref{table:5}. The performance of `$-$Unit Pretraining' is obtained by directly finetuning the mAV-HuBERT on the multilingual video-text without using visual speech units. In this case, the overall VSR performances are dropped especially for non-English languages. Moreover, we require more training times compared to the proposed method as shown in Table \ref{table:3}. When we do not utilize curriculum learning (\ie, `$-$Curriculum'), the performance dramatically decreases and even shows worse performances than the without pre-training method in some languages. This shows that directly performing the visual speech unit to text translation from scratch is challenging to the model in finding the optimal points. Therefore, the proposed curriculum learning is crucial when learning from the visual speech units. These results also coincide with the previous methods utilizing multi-modal complementary in VSR training \cite{afouras2020asr,shi2021learning,zhao2020hearing,ren2021learning,kim2022distinguishing,yeo2023akvsr}. Finally, when we directly perform multilingual VSR with the pre-trained model (\ie, with visual speech units), overall performances are decreased as fine information is lost during the quantization process. Therefore, finetuning with the continuous features for a few epochs should be performed to maximize the pre-trained knowledge.

\begin{table}[t]
\renewcommand{\arraystretch}{1.9}
  \renewcommand{\tabcolsep}{3.5mm}
\centering
\caption{Multilingual VSR performance (WER, \%) comparisons. As there is no prior work that can perform multilingual VSR with a single model, we train AV-HuBERT to perform multilingual VSR.}
  \resizebox{0.9\linewidth}{!}{
  \begin{tabular}{cccccc}
    \Xhline{3\arrayrulewidth}
    \textbf{Method} & \textbf{En} & \textbf{Es} & \textbf{It} & \textbf{Fr} & \textbf{Pt} \\ \hline
    AV-HuBERT \cite{shi2021learning} & \textbf{23.3} & 51.2 & 54.8 & 61.0 & 55.3 \\ \hdashline
    \textbf{Proposed Method} & 24.4 & \textbf{42.7} & \textbf{49.7} & \textbf{55.2} & \textbf{50.6} \\
    \Xhline{3\arrayrulewidth}
  \end{tabular}}
  \label{table:6}
\vspace{-0.3cm}
\end{table}

\subsubsection{Multilingual Visual Speech Recognition with a Single Trained Model}
\label{sec:4.3.5}
We validate the effectiveness of the proposed multilingual VSR method by comparing it with 1) the multilingual VSR model and 2) the monolingual VSR model. Since there is no prior work exploring multilingual VSR with a single model, we train AV-HuBERT \cite{shi2021learning} to perform multilingual VSR and set it as our baseline. Moreover, since the previous non-English VSR methods are language-specific, we compare the performance of our model with multiple monolingual models.

\begin{table*}[t]
  \renewcommand{\arraystretch}{1.3}
  \renewcommand{\tabcolsep}{2.5mm}
\centering
\vspace{-0.1cm}
\caption{VSR performance comparisons with the previous VSR methods. Please note that the proposed method utilizes a single model while the other methods utilize multiple models. Best and second-best scores are bolded and underlined.}
  \resizebox{0.85\linewidth}{!}{
  \begin{tabular}{ccccccc}
    \Xhline{3\arrayrulewidth}
    \textbf{Language} & \textbf{Method} & \makecell{\textbf{Pre-training} \\ \textbf{Data (hrs)}} & \makecell{\textbf{Language-specific} \\ \textbf{Training Data (hrs)}} & \makecell{\textbf{Monolingual} \\ \textbf{Model} } & \makecell{\textbf{Single} \\ \textbf{Multilingual} \\ \textbf{Model} } & \textbf{WER(\%)}  \\ \hline
    \multirow{8}{*}{\textbf{En}} &
    \citet{ma2022visual}  & - & 1,459 & \cmark &  & 31.5  \\
    & \citet{prajwal2022sub} & - & 2,676 & \cmark &  & 30.7  \\
    & AV-HuBERT~\cite{shi2021learning} & 1,759 & 433 & \cmark &  & 28.6  \\
    & VATLM~\cite{zhu2023vatlm} & 1,759 & 433 & \cmark &  & 28.4  \\
    & \citet{haliassos2022jointly} & 1,759 & 433 & \cmark &  & 27.8  \\
    & AKVSR~\cite{yeo2023akvsr} & 1,759 & 433 & \cmark &  & 27.6  \\
    & Auto-AVSR~\cite{ma2023auto} & - & 3,448 & \cmark &  & \textbf{20.5}  \\
    \cdashline{2-7}
    & \textbf{Proposed Method} & 5,512 & 3,258 &  & \cmark & \underline{24.4}  \\
    \hline

    \multirow{4}{*}{\textbf{Es}} & 
    \citet{ma2022visual} & 1,459 & 87 & \cmark &  & 56.3  \\
    & \citet{kim2023lip} & 3,448 & 72 & \cmark &  & 56.9  \\ 
    & \citet{yeo2023visual} & 3,448 & 384 & \cmark &  & \underline{45.7}  \\
    \cdashline{2-7}
    &\textbf{Proposed Method} & 5,512 & 384 &  & \cmark & \textbf{42.7}  \\
    \hline    

    \multirow{4}{*}{\textbf{It}} & 
    \citet{ma2022visual} & 1,459 & 46 & \cmark &  & 57.4  \\
    &\citet{kim2023lip} & 3,448 & 46 & \cmark &  & 59.7  \\ 
    &\citet{yeo2023visual} & 3,448 & 152 & \cmark &  & \underline{51.8} \\
    \cdashline{2-7}
    &\textbf{Proposed Method} & 5,512 & 152 &  & \cmark & \textbf{49.7} \\
    \hline

    \multirow{4}{*}{\textbf{Fr}} & 
    \citet{ma2022visual} & 1,459 & 100 & \cmark & & 66.2  \\
    &\citet{kim2023lip} & 3,448 & 85 & \cmark & & 64.9  \\
    &\citet{yeo2023visual} & 3,448 & 331 & \cmark & & \underline{58.3}  \\
    \cdashline{2-7}
    &\textbf{Proposed Method} & 5,512 & 331 & & \cmark & \textbf{55.2}  \\
    \hline

    \multirow{4}{*}{\textbf{Pt}} & 
    \citet{ma2022visual} & 1,459 & 99 & \cmark &  & 61.5  \\
    &\citet{kim2023lip} & 3,448 & 82 & \cmark &  & 58.6  \\ 
    &\citet{yeo2023visual} & 3,448 & 420 & \cmark &  & \textbf{47.9}  \\  
    \cdashline{2-7}
    &\textbf{Proposed Method} & 5,512 & 420 &  & \cmark & \underline{50.6}  \\
     
    \Xhline{3\arrayrulewidth}
  \end{tabular}}
\label{table:7}
\vspace{-0.05cm}
\end{table*}

\textbf{Comparison with multilingual VSR method.}
Table \ref{table:6} shows the performance comparison results of multilingual VSR methods. Both the AV-HuBERT and the proposed method are finetuned on 4,545 hours of multilingual video-text paired data. The proposed method outperforms the AV-HuBERT for all languages except English. In particular, the proposed method demonstrates significantly improved performance for Es, It, Fr, and Pt, with gains of more than 4\% WERs. For the high-resource language En, the proposed method achieves similar performance with AV-HuBERT but slightly falls behind. Considering multilingualism, the results confirm that the proposed VSR framework using visual speech units is much more effective in building multilingual VSR models by achieving new state-of-the-art performances. It is also worth noting that the proposed method is more efficiently trainable with the proposed pre-training strategy as discussed in Sec. \ref{sec:4.3.2}.

\textbf{Comparison with monolingual VSR method.}
Here, we compare the proposed multilingual VSR performances with the previous state-of-the-art monolingual VSR methods. Please note that the proposed method utilizes a single trained model across the languages, while different methods utilize multiple language-specific VSR models. The results are shown in Table \ref{table:7}. By comparing with the recent state-of-the-art method \cite{yeo2023visual}, we outperform it in 3 languages Spanish (Es), Italian (It), and French (Fr), by 3.0\%, 2.1\%, and 3.1\% WER, respectively. In the English VSR, our method achieves 24.4\% WER while the current previous method \cite{ma2023auto} achieves 20.5\% WER. As discussed in \cite{radford2023whisper}, the performance of high-resource languages can be further improved by scaling the model size to accommodate the diversity of languages. As a result, the proposed multilingual VSR model achieves the best score in Es, It, and Fr, and the second-best score in En and Pt with a single-trained model. Through the comparisons, we can confirm not only the computation efficiency but also the effectiveness of the proposed method in multilingual VSR by outperforming and achieving comparable results with the previous language-specific VSR methods.

\vspace{-0.1cm}
\section{Conclusion}
In this paper, we proposed an efficient multilingual VSR method using a single model. Specifically, we proposed to use visual speech units for pre-training a VSR model to mitigate the huge computational loads in building massive multilingual visual speech modeling. With the proposed strategy, we can greatly reduce the data size and effectively pre-train the VSR model on large-scale multilingual VSR databases. By analyzing the visual speech unit, we validated it contains linguistic information and enables visual speech modeling using discrete inputs. To complement visual speech information with audio, we proposed curriculum learning by gradually increasing the task difficulty. Finally, by finetuning the model on continuous features, we set new state-of-the-art multilingual VSR performances by achieving comparable VSR performances with the previous language-specific VSR models. To the best of our knowledge, this is the first work exploring the multilingual VSR and employing visual speech units as inputs of VSR systems.

\bibliographystyle{ACM-Reference-Format}
\bibliography{sample-base}

\newpage
\appendix
\section{Visualization of Speech Units}
\begin{figure*}[t]
	\centering
	\centerline{\includegraphics[width=14cm]{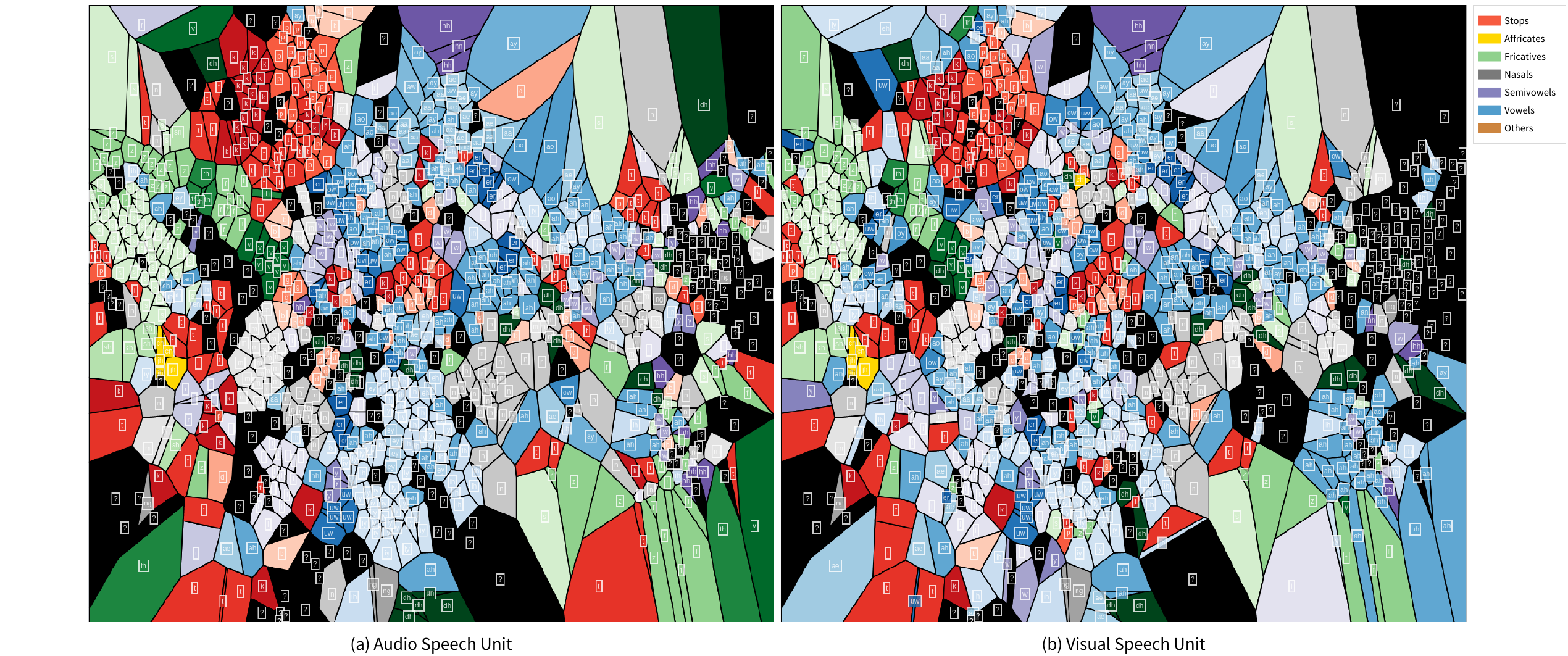}}
    \vspace{-0.4cm}
	\caption{Visualization of speech units. Each boundary represents a single unit and the same color represents the same phoneme or phoneme family. (a) Audio speech unit. (b) Visual speech unit.
	}
	\label{fig:5}
\end{figure*}

Fig. \ref{fig:5} shows the visualization of all 1,000 units of both the audio speech units and the visual speech units. In visual speech units, more units are classified as vowels while audio speech units have more distinct phonemes. As we discussed before, the ambiguity of lip movements is reflected in the figure.

\section{Dataset Statistics for Each Language}
The dataset statistics for each language are shown in Table \ref{table:8}. Please note that there are only 181,034 non-English human-labeled videos (\ie, mTEDx). Therefore, we increase the quantity of labeled data by utilizing the automatic labels proposed by \cite{yeo2023visual, ma2023auto}. With this, we can construct 2,014,212 multilingual video-text paired data. The number of these automatic labels can be found in the `Auto-labeled \# of Video' column in the table. 
\begin{table}[h]
	\renewcommand{\arraystretch}{1.2}
	\renewcommand{\tabcolsep}{1.5mm}
\centering
\resizebox{0.999\linewidth}{!}{
\begin{tabular}{cccccc}
\Xhline{3\arrayrulewidth}
\textbf{Language} &\textbf{Dataset} & \textbf{\makecell{Human-labeled \\ \# of Video}} & \textbf{\makecell{Auto-labeled \\ \# of Video}} & \textbf{Hours} &\textbf{\makecell{Total\\Hours}} \\ \hline

\multirow{4}{*}{\textbf{En}} 
& LRS2 & 142,157  & - & 223 & \multirow{4}{*}{3,481}  \\
&LRS3 & 150,498  & - & 433 & \\
& VoxCeleb2 & -  & 628,418 & 1,326 & \\ 
& AVSpeech & -  & 837,044 & 1,499 & \\ 
\hline

\multirow{3}{*}{\textbf{Es}} 
& mTEDx & 44,532  & - & 72 & \multirow{3}{*}{384}  \\
& VoxCeleb2 & -  & 22,682 & 42 & \\
& AVSpeech & -  & 151,173 & 270 & \\ 
\hline

\multirow{3}{*}{\textbf{It}} 
& mTEDx & 26,018  & - & 46 & \multirow{3}{*}{152}  \\
& VoxCeleb2 & -  & 19,261 & 38 & \\
& AVSpeech & -  & 38,227 & 68 & \\ 
\hline

\multirow{3}{*}{\textbf{Fr}} 
& mTEDx & 58,426  & - & 85 & \multirow{3}{*}{331}  \\
& VoxCeleb2 & -  & 66,943 & 124 & \\
& AVSpeech & -  & 69,020 & 122 & \\ 
\hline

\multirow{3}{*}{\textbf{Pt}} 
& mTEDx & 52,058  & - & 82 & \multirow{3}{*}{420}  \\
& VoxCeleb2 & -  & 4,843 & 9 & \\
& AVSpeech & -  & 176,601 & 329 & \\ 
\hline

\multirow{2}{*}{\textbf{De}} 
& VoxCeleb2 & -  & - & 190 & \multirow{2}{*}{333}  \\
& AVSpeech & -  & - & 143 & \\
\hline

\multirow{2}{*}{\textbf{Ru}} 
& VoxCeleb2 & - & - & 2 & \multirow{2}{*}{288}  \\
& AVSpeech & -  & - & 286 & \\
\hline

\multirow{2}{*}{\textbf{Ar}} 
& VoxCeleb2 & -  & - & 7 & \multirow{2}{*}{114}  \\
& AVSpeech & -  & - & 107 & \\
\hline

\multirow{2}{*}{\textbf{El}} 
& VoxCeleb2 & -  & - & 1 & \multirow{2}{*}{9}  \\
& AVSpeech & -  & - & 8 & \\

\Xhline{3\arrayrulewidth}
\end{tabular}}
\caption{Data statistics of each language used in this work including automatic labels.}
\label{table:8}
\end{table}

\section{Analysis of Linguistic Information in Visual Speech Units}
To evaluate the extent to which linguistic information is maintained in visual speech units compared to continuous representations, we assess the speech recognition performances (WER) using five different input representations: raw audio, audio features, visual features, audio speech units, and visual speech units, on the LRS3 dataset. The results are shown in Table \ref{table:9}. In particular, we can confirm that discretization does not significantly compromise the linguistic content of the visual speech, with WERs ranging from 38.1\% to 41.9\%. Similarly, the audio speech units also demonstrate sufficient performance in retaining the linguistic content. By comparing these results with the speaker verification experiments in Table \ref{table:4}, we can confirm that the visual speech units contain linguistic information more than the speaker information.

\begin{table}[h]
\renewcommand{\arraystretch}{1.5}
\renewcommand{\tabcolsep}{1.0mm}
\centering
\resizebox{0.99\linewidth}{!}{
  \begin{tabular}{ccccc}
  \hline
  \textbf{Raw audio} & \textbf{Audio Feature} & \textbf{Visual Feature} & \textbf{Audio Speech unit} & \textbf{Visual speech unit} \\ \hline
  2.7\%  & 4.2\% & 38.1\% & 11.5\% & 41.9\% \\ \hline
  \end{tabular}}
\caption{WER comparisons using different inputs.}
\vspace{-0.4cm}
\label{table:9}
\end{table}

\begin{table}[h]
\renewcommand{\arraystretch}{1.2}
  \centering
  \resizebox{0.7\linewidth}{!}{
  \begin{tabular}{cccccc}
  \hline
    \textbf{Method} & \textbf{En} & \textbf{Es} & \textbf{It}  & \textbf{Fr} & \textbf{Pt} \\ \hline
    AV $\rightarrow$ V & 39.4\%  & 69.5\% & 72.9\% & 78.0\% & 77.0\% \\
    Curriculum & 37.5\% & 68.1\% & 70.3\% & 76.3\% & 74.0\% \\ \hline
\end{tabular}}
\caption{Performance of different pre-training strategies.}
\vspace{-0.4cm}
\label{table:10}
\end{table}

\section{Effectiveness of Curriculum Learning}
Instead of using curriculum learning, which starts the training with audio-visual speech units and gradually masks out the audio stream part, another training strategy could be transfer learning. This consists of two stages: pre-training the model with audio-visual speech units and then transferring to visual-only speech units. Both strategies share the objective of initially training the model with rich audio-visual speech information and then transforming the model to work with visual-only speech. To evaluate the effectiveness of these two strategies, we report the performances of 1) Transfer Learning, which involves the two-stage learning process, and 2) Curriculum Learning, which uses the masking strategy. The results are shown in Table \ref{table:10}. The performances of both methods are measured without finetuning on the continuous representations. By comparing these results, we confirm that the curriculum learning strategy showed relative improvements over the sequential training approach (\ie, transfer learning).

\section{Detailed Training Setup}
We provide the detailed training setup used for experiments in Table \ref{table:11}. For pre-training mAV-HuBERT, we use a polynomial decay Learning Rate (LR) scheduler, batch size of 1,000 frames for each GPU, and train steps of 350k. For pre-training with the visual speech unit, we use 3,000 frames per GPU even though we can increase it to 6,000 frames. During finetuning, the pre-trained encoder is frozen for 10k steps and 7.2k steps for multilingual finetuning.
\begin{table*}[t]
	\renewcommand{\arraystretch}{1.1}
	\renewcommand{\tabcolsep}{2mm}
\centering
\vspace{0.1cm}
\resizebox{0.65\linewidth}{!}{
\begin{tabular}{ccccc}
\Xhline{3\arrayrulewidth}
& \makecell{\textbf{Pre-training} \\ \textbf{(mAV-HuBERT)}} & \makecell{\textbf{Pre-training} \\ \textbf{(Visual speech unit to} \\ \textbf{text translation)}} & \makecell{\textbf{Fine-tuning} \\ \textbf{(Multilingual)}} & \makecell{\textbf{Fine-tuning} \\   \textbf{(Monolingual)}}  \\ \hline
\# of epochs & 40 & 11 & 8 & -\\
\# of steps & 350,000 & 60,000 & 120,000 & 60,000\\
\# of frozen steps & - & - & 10,000 & 7,200 \\
\# of GPUs & 64 & 32 & 32 & 8 \\
Max frames / batch & 1000 & 3000 & 1000 & 1000 \\
LR scheduler & polynomial decay & tri-stage & tri-stage & tri-stage \\
warmup updates & 48,000 & 15,000 & 15,000 & 15,000 \\
peak learning rate & 2e-3 & 1e-3 & 4e-4 & 4e-4   \\
Adam $(\beta_{1}, \beta_{2})$ & (0.9, 0.98) & (0.9, 0.98) & (0.9, 0.98) & (0.9, 0.98)  \\
\Xhline{3\arrayrulewidth}
\end{tabular}}
\caption{Details of hyperparameters used in training.}
\label{table:11}
\end{table*}


\section{Examples of Predicted Sentences}
\begin{figure*}[t]
	\centering
    \vspace{-0.3cm}
	\centerline{\includegraphics[width=12cm]{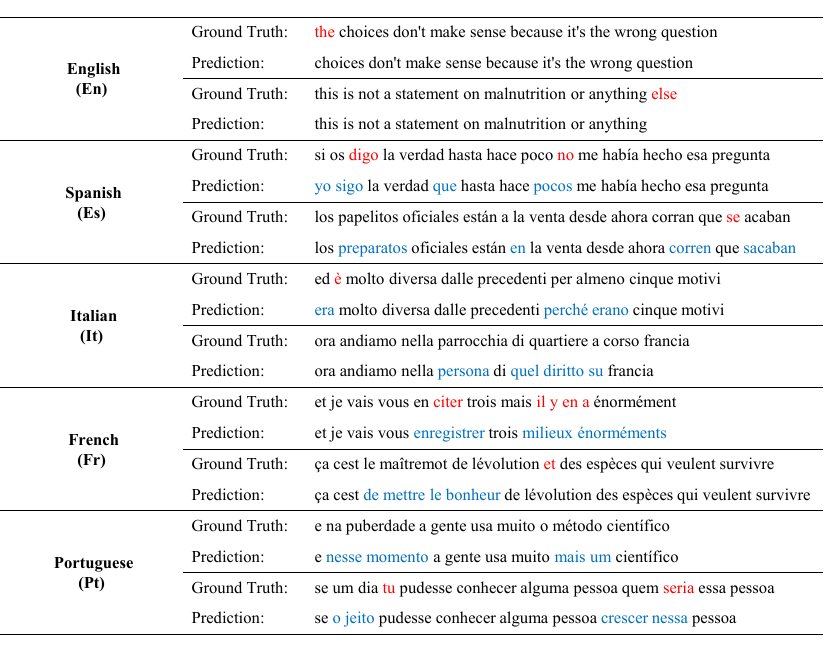}}
    \vspace{-0.4cm}
	\caption{Example sentences predicted from the single proposed multilingual VSR model on LRS3 and mTEDx test set. The \textcolor[rgb]{0.9,0,0}{Red} and \textcolor[rgb]{0,0.2,0.90}{Blue} indicate deletion and wrong predicted words, respectively.}
	\label{fig:6}
    \vspace{-0.4cm}
\end{figure*}

We show some examples of predicted transcriptions by the proposed multilingual VSR model and ground-truth transcriptions in Fig. \ref{fig:6}. For each language, we show two examples. The red-colored words indicate the deletion error and the blue-colored words indicate the wrong prediction (\ie, insertion or substitution)

\end{document}